\documentclass[runningheads]{llncs}

\usepackage{hyperref}
\usepackage{amsmath}
\usepackage{amssymb}
\usepackage{graphicx}
\usepackage{algorithm}
\usepackage{algorithmic}
\usepackage{subfigure}
\usepackage{threeparttable}

\begin{document}
  \pagestyle{headings}


  \title{Mining top-k granular association rules for recommendation}

  \author{Fan Min \and William Zhu}

  \institute{Lab of Granular Computing,\\ Minnan Normal University, Zhangzhou 363000, China\\
  \email{minfanphd@163.com, williamfengzhu@gmail.com}}
  \maketitle

  \begin{abstract}
Recommender systems are important for e-commerce companies as well as researchers.
Recently, granular association rules have been proposed for cold-start recommendation.
However, existing approaches reserve only globally strong rules; therefore some users may receive no recommendation at all.
In this paper, we propose to mine the top-k granular association rules for each user.
First we define three measures of granular association rules.
These are the source coverage which measures the user granule size, the target coverage which measures the item granule size, and the confidence which measures the strength of the association.
With the confidence measure, rules can be ranked according to their strength.
Then we propose algorithms for training the recommender and suggesting items to each user.
Experimental are undertaken on a publicly available data set MovieLens.
Results indicate that the appropriate setting of granule can avoid over-fitting and at the same time, help obtaining high recommending accuracy.
    \end{abstract}

  %
  %
  \section{Introduction}\label{section: introduction}
The quality of the recommender system is essential to an e-commerce company \cite{SarwarB2000Analysis}.
Due to the huge data and the uncertainty of user behavior, recommender systems have also attracted much research interests from the academia \cite{Adomavicius2005Toward}.
Tapestry \cite{Goldberg1992usingcollaborative} and GroupLens \cite{Resnick1994Grouplens} are among the earliest implementations of collaborative filtering based recommender systems.
Many content-based filtering methods (see, e.g., \cite{Foltz1992Personlized,Mooney2000Content}) have been developed.
Currently, people tend to adopt more sophisticated ensemble methods (see, e.g., \cite{Jahrer10Combining,Bothos11Information}) that aggregate predications of numerous base algorithms.

Cold-start recommendation \cite{Cremonesi11Hybrid,ScheinA2002ColdStart} is a challenging problem in recommender systems.
It refers the situation that a new user or a new item has just entered the system.
Researchers have addressed the problem mainly through content-based filtering methods (see, e.g., \cite{Mooney2000Content,PazzaniM2007Content}).
Recently, inspired by granular computing \cite{YaoVasilakosPedrycz2013Granular,YaoDeng2013Paradigm,ZhuWang03Reduction}, granular association rules with four measures have also bee applied to the problem.
An example of such a rule might be ``42\% young women rate at least 33\% adventure movies released in 1990s; 21\% users are young women and 15\% movies are adventure ones released in 1990s."
Here 21\%, 15\%, 42\% and 33\% are the source coverage, the target coverage, the source confidence, and the target confidence, respectively.
However, the existing rule mining problem requires four thresholds, one for each measure.
Consequently, only globally strong rules are reserved, some users may receive many recommendations, and some may receive no recommendation at all.

In this paper, we propose to mine the top-k granular association rules for each user.
First, we define granular association rules with three measures.
An example of such a rule might be ``young women rate adventure movies released in 1990s with a probability of 35\%; 21\% users are young women and 15\% movies are adventure ones released in 1990s."
Here 35\% is the confidence of the rule.
With the confidence measure, the strength of any two rules can be compared.
In fact, the four measure suit and the three measure suit are appropriate for different situations.
One cannot compute the confidence according to the source confidence and the target confidence, or vice versa.
Then, we propose an algorithm for training the recommender.
This is done through building connections between source and target granules that satisfy the coverage thresholds.
Finally we propose an algorithm to obtain top-k rules which in turn suggest k types of items to each user.

The duplicate recommendation issue may arise in the rule matching process.
For example, we have two rules ``young women rate adventure movies released in 1990s with a probability of 35\%" and ``young students rate adventure movies released in 1990s with a probability of 38\%."
If a new user is a young female student, she will match both rules that produce the same recommendation, namely, movies released in 1990s.
In this case, we discard the weaker rule, and the confidence will be viewed 38\%.

Experiments are undertaken on the MovieLens data set \cite{movielens} using our open source software Grale \cite{Grale}.
We tested different thresholds and different k values with a number of observations.
First, with some appropriate settings, the recommending accuracy is about 29\% for the new user problem, compared with only 6.2\% for random recommendation.
Second, with the increase of k value, the confidence of recommendation decreases smoothly.
Third, when the coverage thresholds are lower, the recommender performs better on the training set however worse on the testing set.
This is due to the over-fitting of the recommender.
On the other hand, when the coverage thresholds are high, the recommender performs similar on both sets.
Therefore, the appropriate setting of granule is essential to the performance of the recommender.

  %
  %
  \section{Preliminaries}\label{section: preliminaries}
In this section, we review some preliminary knowledge such as many-to-many entity-relationship systems and information granules \cite{MinHuZhu12GranularFour,MinHuZhu13GranularTwo}.
We also propose granular association rules with three measures.

  %
  %
  \subsection{Many-to-many entity-relationship systems}\label{subsection: scaled-positive}
First we revisit the definitions of information systems, binary relations and many-to-many entity relationship systems \cite{MinHuZhu12GranularFour}.
\begin{definition}\label{definition: ins}
$S = (U, A)$ is an information system, where $U = \{x_1, x_2, \dots, x_n\}$ is the set of all objects, $A = \{a_1, a_2, \dots, a_m\}$ is the set of all attributes, and $a_j(x_i)$ is the value of $x_i$ on attribute $a_j$ for $i \in [1..n]$ and $j \in [1..m]$.
\end{definition}

Two information systems are listed in Tables \ref{subtable: user} and \ref{subtable: movie}, respectively.
In Table \ref{subtable: movie}, 1 indicates \emph{true}, and 0 indicates \emph{false}.

\begin{definition}\label{definition: binary-relation}
Let $U = \{x_1, x_2, \dots, x_n\}$ and $V = \{y_1, y_2, \dots, y_k\}$ be two sets of objects.
Any $R \subseteq U \times V$ is a binary relation from $U$ to $V$.
\end{definition}

An example of binary relation is given by Table \ref{subtable: rates}, where $U$ is the set of users as indicated by Table \ref{subtable: user}, and $V$ is the set of movies as indicated by Table \ref{subtable: movie}.
A binary relation can be viewed as an information system.
However, in order to save space, it is more often stored in the database as a table with two foreign keys.

\begin{definition}\label{definition: m-m-er}
A many-to-many entity-relationship system (MMER) is a 5-tuple $ES = (U, A, V, B, R)$, where $(U, A)$ and $(V, B)$ are two information systems, and $R \subseteq U \times V$ is a binary relation from $U$ to $V$.
\end{definition}
An example of MMER is given by Table \ref{table: mmer}.

\begin{table}[tb]\caption{A many-to-many entity-relationship system}\label{table: mmer}
\centering
\setlength{\tabcolsep}{19pt}
\subtable[User]{
\begin{tabular*}{12cm}{@{\extracolsep{\fill}}ccccc}
\hline
User-id     & Age            & Gender &  Occupation \\
\hline
1           & $[18, 24]$     & M      &  technician \\
2           & $[50, 55]$     & F      &  other      \\
3           & $[18, 24]$     & M      &  writer     \\
\dots       \\
943         & $[18, 24]$     & M      &  student    \\
\hline
\end{tabular*}
\label{subtable: user}
}
\qquad
\setlength{\tabcolsep}{2pt}
\subtable[Movie]{
\begin{tabular*}{12cm}{@{\extracolsep{\fill}}cccccccccc}
\hline
Movie-id       & Release-decade &  Action & Adventure & Animation & \dots & Western \\
\hline
1        & 1990s          &  0      & 0         & 0         & \dots & 0\\
2        & 1990s          &  0      & 1         & 1         & \dots & 0\\
3        & 1990s          &  0      & 0         & 0         & \dots & 0\\
\dots    & \\
1,682    & 1990s          &  0      & 0         & 0         & \dots & 0\\
\hline
\end{tabular*}
\label{subtable: movie}
}
\qquad
\setlength{\tabcolsep}{2pt}
\subtable[Rates]{
\begin{tabular*}{12cm}{@{\extracolsep{\fill}}cccccccc}
\hline
User-id$\diagdown$ Movie-id &  1     &  2     &  3    & 4      & 5      & \dots & 1,682\\
\hline
1      &  0      &  1      &  0     &  1      & 0     & \dots  & 0\\
2      &  1      &  0      &  0     &  1      & 0     & \dots  & 1\\
3      &  0      &  0      &  0     &  0      & 1     & \dots  & 1\\
\dots  & \\
943    &  0      &  0      &  1     &  1      & 0     & \dots  & 1\\
\hline
\end{tabular*}
\label{subtable: rates}
}
\end{table}

  %
  %
  \subsection{Information granules}\label{subsection: granules}
Users and items can be described by information granules \cite{MinHuZhu13GranularTwo,YaoDeng2013Paradigm}.
In an information system, any $A' \subseteq A$ induces an equivalence relation \cite{Pawlak82Rough,SkowronStepaniuk94Approximation}
\begin{equation}\label{equation: equivalence-relation}
E_{A'} = \{(x, y) \in U \times U| \forall a \in A', a(x) = a(y)\},
\end{equation}
and partitions $U$ into a number of disjoint subsets called \emph{blocks} or \emph{granules}.
The granule containing $x \in U$ is
\begin{equation}\label{equation: block-contain-x}
E_{A'}(x) = \{y \in U| \forall a \in A', a(y) = a(x)\}.
\end{equation}

The following definition was employed by Yao and Deng \cite{YaoDeng2013Paradigm}.
\begin{definition}\label{definition: granule}
A granule is a triple
\begin{equation}\label{equation: granule}
G = (g, i(g), e(g)),
\end{equation}
where $g$ is the name assigned to the granule, $i(g)$ is a representation of the granule,
and $e(g)$ is a set of objects that are instances of the granule.
\end{definition}

According to Equation (\ref{equation: block-contain-x}), $(A', x)$ determines a granule in an information system.
Hence $g = g(A', x)$ is a natural name to the granule.
$i(g)$ can be formalized as the conjunction of respective attribute-value pairs, i.e.,
\begin{equation}\label{equation: intension-granule}
i(g(A', x)) =  \bigwedge_{a \in A'}\langle a: a(x) \rangle.
\end{equation}
$e(g)$ is given by
\begin{equation}\label{equation: extension-granule}
e(g(A', x)) = E_{A'}(x).
\end{equation}

Let $x \in U$ and $A'' \subset A' \subseteq A$, we have
\begin{equation}\label{equation: finer-granule}
e(g(A', x)) \subseteq e(g(A'', x)).
\end{equation}
Consequently, we say that $g(A', x)$ is \emph{finer} than $g(A'', x)$, and $g(A'', x)$ is \emph{coarser} than $g(A', x)$.

  %
  %
  \subsection{Granular association rules with three measures}\label{subsection: grarule-rules}
Now we discuss the means for connecting users and items.
A \emph{granular association rule} \cite{MinHuZhu12GranularFour,MinHuZhu13GranularTwo} is an implication of the form
\begin{equation}\label{equation: granular-association}
(GR): \bigwedge_{a \in A'}\langle a: a(x) \rangle \Rightarrow \bigwedge_{b \in B'}\langle b: b(y) \rangle,
\end{equation}
where $A' \subseteq A$ and $B' \subseteq B$.

Before defining evaluation measures, let us look at an example granular association rule ``young women rate adventure movies released in 1990s with a probability of 35\%; 21\% users are young women and 15\% movies are adventure ones released in 1990s."
Here 21\%, 15\%, and 35\% are the source coverage, the target coverage, and the confidence, respectively.

According to Equations (\ref{equation: intension-granule}) and (\ref{equation: extension-granule}), the set of objects meeting the left-hand side of the granular association rule is
\begin{equation}\label{equation: left-granular-rule}
LH(GR) = E_{A'}(x);
\end{equation}
while the set of objects meeting the right-hand side of the granular association rule is
\begin{equation}\label{equation: right-granular-rule}
RH(GR) = E_{B'}(y).
\end{equation}

The \emph{source coverage} of $GR$ is
\begin{equation}\label{equation: source-coverage}
scov(GR) = \frac{|LH(GR)|}{|U|};
\end{equation}
while the \emph{target coverage} of $GR$ is
\begin{equation}\label{equation: target-coverage}
tcov(GR) = \frac{|RH(GR)|}{|V|}.
\end{equation}

The \emph{confidence} of $GR$ is the probability that a user chooses an item, namely
\begin{equation}\label{equation: confidence}
conf(GR) = \frac{|(LH(GR) \times RH(GR)) \cap R|}{|LH(GR)| \times |RH(GR)|}.
\end{equation}

The confidence measure is a new concept of this work.
It enables the comparison of the strength of any two rules.
Therefore it is a key concept supporting top-k rule mining.
Note that in existing works, the \emph{source confidence} and the \emph{target confidence} measures were defined.
Here we do not revisit their definitions to avoid confusion.

  %
  %
  \section{Mining top-k granular association rules}\label{section: rule-mining}
Generally, a recommender system should recommend a number of items to each user.
In order to make more successful recommendations, we would like to match each user to granular association rules, and choose strong rules for item suggestion.
At the same time, the number of recommendations might be controlled approximately the same.
Therefore we propose to mine top-k granular association rules for each user.

We divide the whole process into the rule set constructing stage and the recommending stage.
In the first stage, a rule set is constructed as indicated by Algorithm \ref{algorithm: rule-set}.
To store the rule set, we need an array to store $SG(ms)$, an array to store $TG(mt)$, and a matrix with size $|SG(ms)| \times |TG(mt)|$ to store confidence of all rules.
Here $SG(ms)$ ($GT(ms)$) is the set of all source (target) granules satisfying the source (target) coverage threshold.

\begin{algorithm}[tb!]\caption{Rule set construction}\label{algorithm: rule-set}
  \textbf{Input}: The training set $ES = (U, A, V, B, R)$, $ms$, $mt$.\\
  \textbf{Output}: Source granules, target granules, and the rule set, all stored in the memory.\\
  \textbf{Method}: training\\
  \begin{algorithmic}[1]
    \STATE $SG(ms) = \{(A', x) \in 2^A \times U| \frac{|E_{A'}(x)|}{|U|} \geq ms\}$;
    \STATE $TG(mt) = \{(B', y) \in 2^B \times V| \frac{|E_{B'}(y)|}{|V|} \geq mt\}$;
    \FOR {each $g \in SG(ms)$}
      \FOR {each $g' \in TG(mt)$}
        \STATE $GR = (i(g) \Rightarrow i(g'))$;
        \STATE compute $conf(GR)$;
      \ENDFOR
    \ENDFOR
  \end{algorithmic}
\end{algorithm}

In the second stage, top-k rules are mined for each user.
Respective item granules are recommended to the user.
The process is described in Algorithm \ref{algorithm: top-k-rule-recommendation}.
The algorithm is exhaustive, hence the top-k rules are always obtained for recommendation.

\begin{algorithm}[tb!]\caption{Top-k rules based recommendation}\label{algorithm: top-k-rule-recommendation}
  \textbf{Input}: The testing set $ES' = (U', A, V', B, R')$.\\
  \textbf{Output}: Recommendation for each object in $U'$.\\
  \textbf{Method}: recommend\\
  \begin{algorithmic}[1]
    \FOR {each $x \in U'$}
      \FOR {each $g \in SG(ms)$}
        \IF {$x$ matches $g$}
          \FOR {each $g' \in TG(mt)$}
            \IF {$g'$ is already among the top-k recommended granules}
                \STATE reserve the higher confidence value;
                \STATE continue;
            \ENDIF
            \STATE $GR = (i(g) \Rightarrow i(g'))$;
            \STATE compare $GR$ with other rules matching $x$, and reserve top-k recommended granules according to the confidence;
          \ENDFOR
        \ENDIF
      \ENDFOR
      \STATE recommend the top-k granules to $x$;
    \ENDFOR
  \end{algorithmic}
\end{algorithm}

Sometimes two rules matching one user may have the same target granule.
For example, we have two rules ``young women rate adventure movies released in 1990s with a probability of 35\%" and ``young students rate adventure movies released in 1990s with a probability of 38\%."
If a new user is a young female student, she will match both rules that produce the same recommendation.
To deal with this situation, we discard the weaker rule, and keep the higher probability, namely 38\% for this example.
In this way, the top-k rules suggest different target granules.
It should be noted further that different target granules may still overlap.
That is natural and we will not try to avoid it.

The performance of the recommender is evaluated mainly by the accuracy of the recommendation.
Formally, let the number of recommended items be $M$, and the number of success recommendations be $N$, the accuracy is $N / M$.

We will compare three training and testing scenarios.
\begin{enumerate}
\item{Random recommendation. There is no training stage. An item is randomly recommended to a user. This is a baseline approach since a recommender which is not significantly better than a random one is simply useless.}
\item{Testing on the training set. It is interesting to test the rule set on the training set.
    Since only attribute values can be employed, the recommending accuracy may not be high.}
\item{Divide the user set into the training and testing set. This scenario corresponds to the new user cold-start problem.}
\item{Divide the item set into the training and testing set. This scenario corresponds to the new item cold-start problem.}
\item{Divide both the user and the item sets. In this scenario, both users and items are new. Hence the problem is very challenging.}
\end{enumerate}

  %
  %
  \section{Experiments}\label{section: experiments}
In this section, we try to answer the following questions through experimentation.

\begin{enumerate}
\item{What is the performance of the recommender for the new user, new item, and both new cold-start problems?}
\item{How does the performance change for different k values?}
\item{How does the number of recommendations change with respect to the granule size?}
\item{How does the performance vary for the training/testing sets?}
\item{How does the performance change for different sampling of the training set?}
\end{enumerate}

  %
  %
  \subsection{Dataset}\label{subsection: data set}
We tested granular association rules on the MovieLens \cite{movielens} which is widely used in recommender systems (see, e.g., \cite{Cremonesi11Hybrid,ScheinA2002ColdStart}).
The database schema is as follows.
\begin{enumerate}
\item[$\bullet$]{User (\underline{userID}, age, gender, occupation)}
\item[$\bullet$]{Movie (\underline{movieID}, release-year, genre)}
\item[$\bullet$]{Rates (\underline{userID, movieID})}
\end{enumerate}
We use the version with 943 users and 1,682 movies.
The data are preprocessed to cope with Definition \ref{definition: m-m-er} as follows.
The original Rate relation contains the rating of movies with 5 scales, while we only consider whether or not a user has rated a movie.
The user age is discretized to 9 intervals as indicated by the data set.
Since there are few movies before 1970s and too many movies after 1990, the release year is discretized to 3 intervals: before 1970s, 1970s-1980s, and 1990s.
The genre is a multi-valued attribute.
Therefore we scale it to 18 boolean attributes and deal with it using the approach proposed in \cite{MinZhu13Multivalued}.

  %
  %
  \subsection{Results}\label{subsection: results}
\begin{figure}[h]
    \centering
        \includegraphics[width=3.5in]{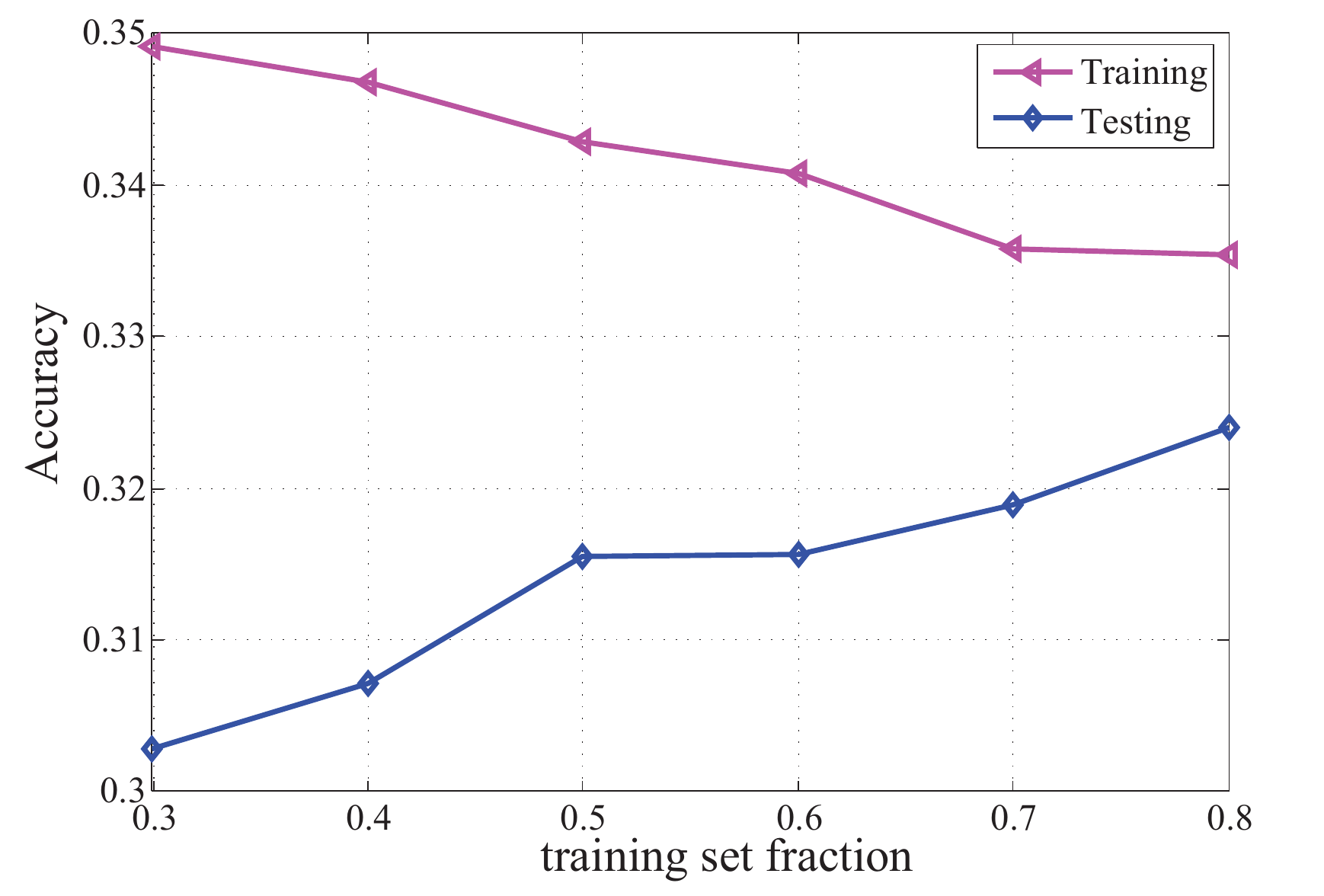}
\caption{Accuracy for different training set fraction}
\label{figure: fraction}
\end{figure}

We undertake five sets of experiments to answer the questions raised at the beginning of the section one by one.
The settings are as follows: the training set fraction is 0.6 except for Fig. \ref{figure: fraction}.
Each experiment is repeated 20 times with different sampling of training and testing sets, and the average accuracy is computed.

\begin{figure}[h]
    \centering
        \includegraphics[width=3.5in]{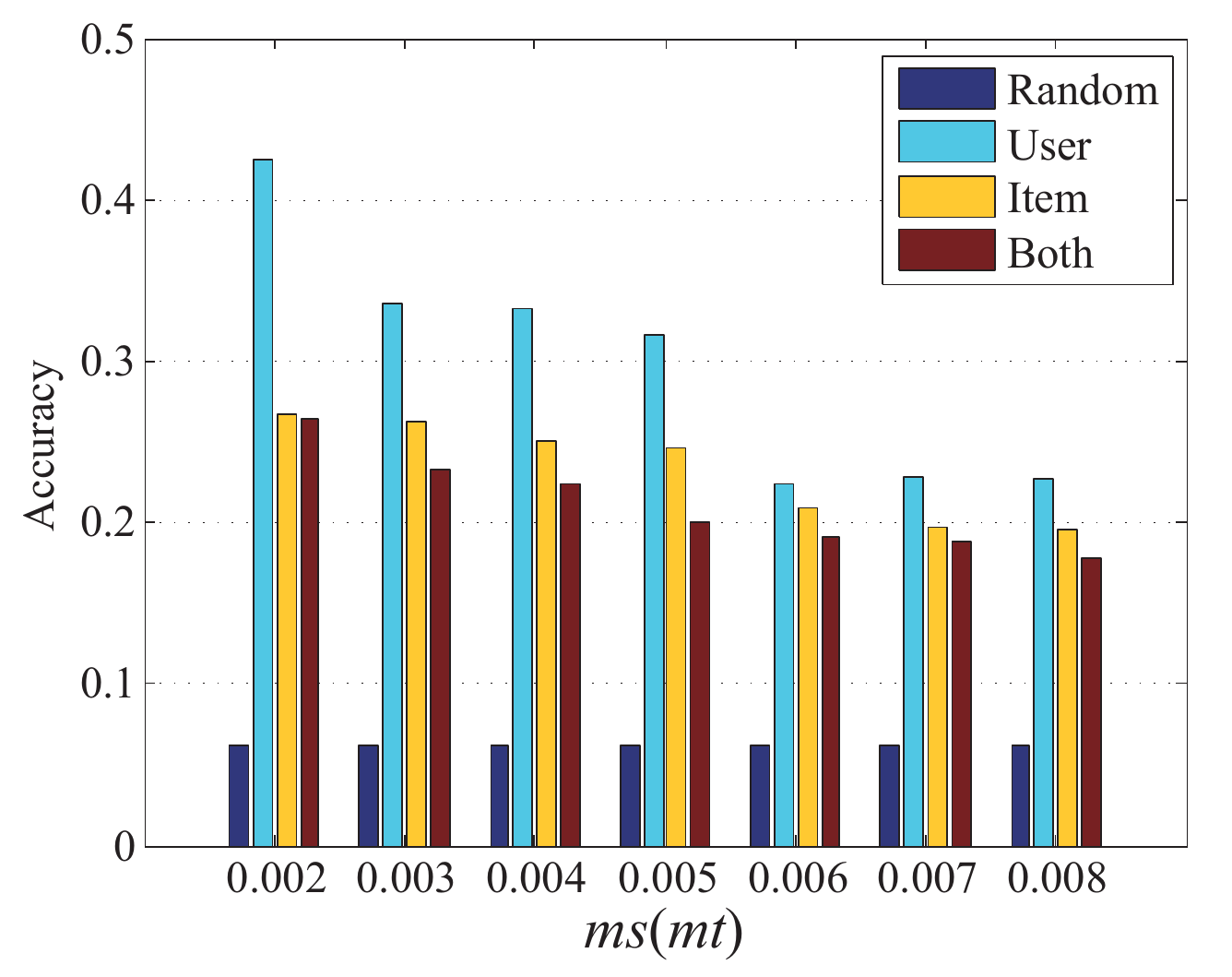}
\caption{Comparison of different scenarios}
\label{figure: scenarios}
\end{figure}

Fig. \ref{figure: scenarios} shows the accuracy of the recommender on the new user, new item, and both new scenarios.
We let k = 1 to test the performance of top-1 rule for each user.
The random recommender, which has an accuracy close to 0.062, is also illustrated for comparison.
The result indicates that the new user problem is easier and respective recommendations are more meaningful.
The both new problem is the hardest since the least information is available.
Moreover, the recommender performs better for smaller granules.

\begin{figure}[h]
    \centering
        \includegraphics[width=3in]{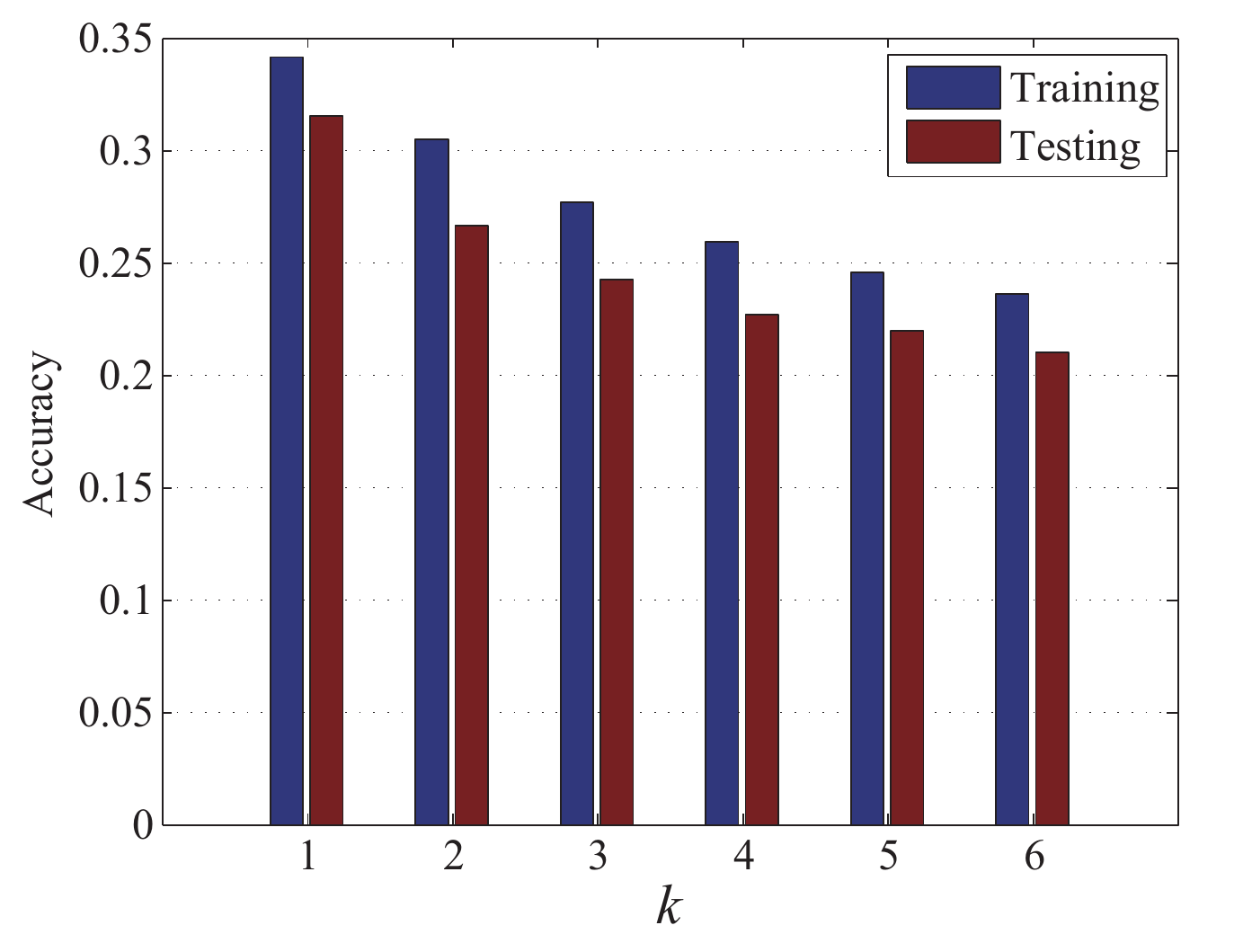}
\caption{Performance of the recommender for different k values}
\label{figure: k}
\end{figure}

Fig. \ref{figure: k} indicates that with the increase of k, the accuracy of the recommender decreases smoothly.
It also shows that the first recommendation is significantly better than the second recommendation, and so on.

\begin{figure}[h]
    \centering
        \includegraphics[width=3.5in]{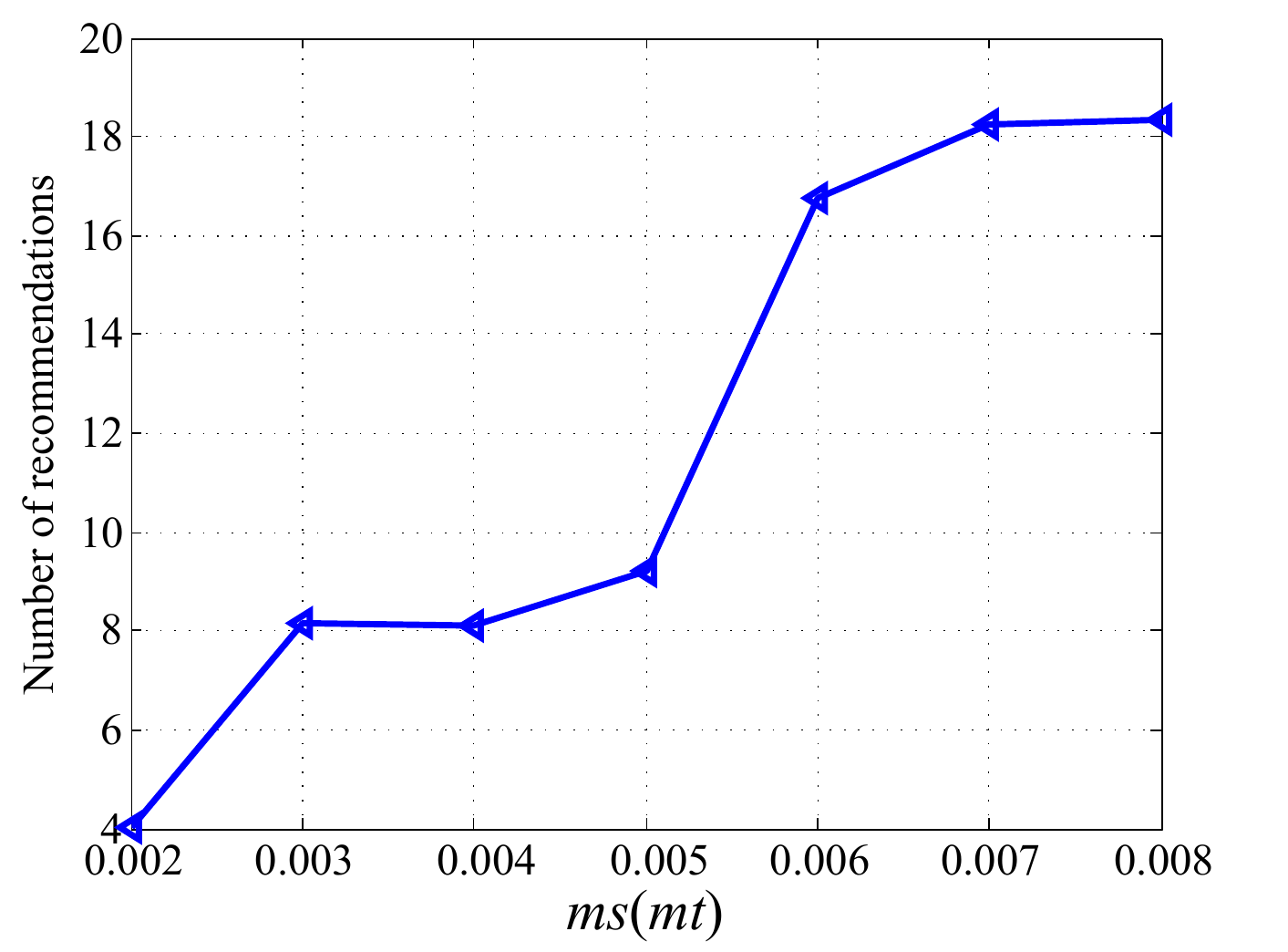}
\caption{Comparison of different granules}
\label{figure: granule}
\end{figure}

Fig. \ref{figure: granule} indicates that with the increase of the granule size, the number of recommended items change at certain thresholds.
This is because that there may not exist granules to exactly match certain given size.

\begin{figure}[h]
    \centering
        \includegraphics[width=3.5in]{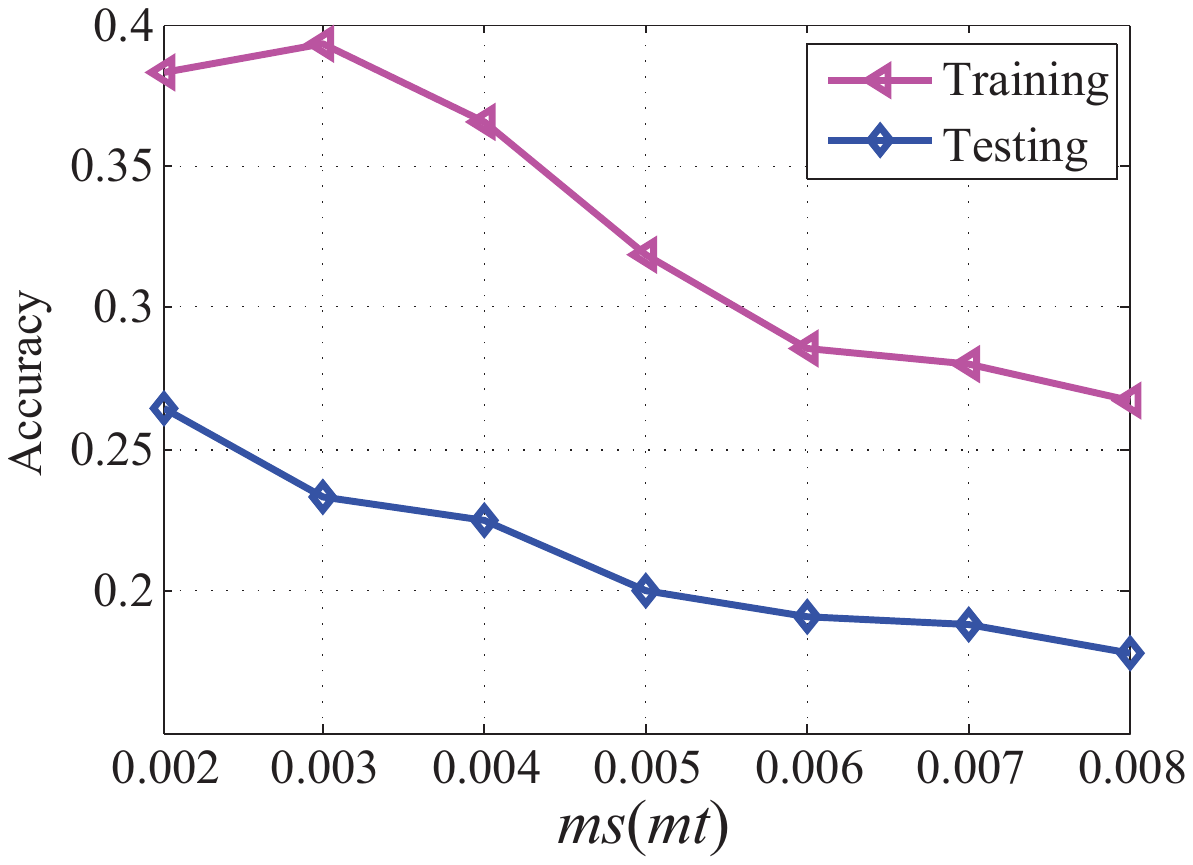}
\caption{Accuracy for training and testing sets}
\label{figure: tvst}
\end{figure}

Fig. \ref{figure: tvst} indicates that the accuracy on the training and testing sets have similar trends.
The trend on the different k values is also validated through Fig. \ref{figure: k}.

Fig. \ref{figure: fraction} indicates that with the increase of the training set and the decrease of the testing set, the accuracy on both sets tend to be closer.
  %
  %
  \section{Conclusions}\label{section: conclusion}
In applications, it is much more easier to specify the number of rules than the support threshold.
Therefore building recommenders based on top-k granular association rules is quite natural.
Experimental results indicate that the appropriate selection of the granule are essential to the performance of the recommender.
In the further, we will design algorithms for large data sets and for effective granule selection.

  %
  %
  \section*{Acknowledgements}\label{section: acknowledgements}
This work is in part supported by National Science Foundation of China under Grant No. 61170128, Fujian Province Foundation of Higher Education under Grant No. JK2012028, and the Natural Science Foundation of Fujian Province, China under Grant No. 2012J01294, and State key laboratory of management and control for complex systems open project under Grant No. 20110106.

  %
  %

\end{document}